\documentclass[aps,prb,groupedaddress,twocolumn]{revtex4}
\usepackage{graphicx}
\usepackage{multirow}

\usepackage{filecontents}

\begin{document}
\title{Random gauge models of the superconductor-insulator transition in two-dimensional disordered superconductors.}

\author{Enzo Granato}

\address{Laborat\'orio Associado de Sensores e Materiais,
Instituto Nacional de Pesquisas Espaciais, 12227-010 S\~ao Jos\'e dos
Campos, SP, Brazil}

\begin{abstract}
We study numerically the superconductor-insulator transition in two-dimensional inhomogeneous superconductors with
gauge disorder, described by four different quantum rotor models: a gauge glass, a flux glass, a binary phase glass  and a Gaussian phase glass. 
The first two models, describe the combined effect of geometrical disorder in the array of local superconducting islands
and a uniform external magnetic field while the last two describe the effects of random negative Josephson-junction couplings or $\pi$ junctions. 
Monte Carlo simulations in the path-integral representation of the models are used to determine the 
critical exponents and the universal conductivity at the quantum phase transition.  The gauge and flux glass models display the same
critical behavior, within the estimated numerical uncertainties. Similar agreement is found for the binary and Gaussian phase-glass models. 
Despite the different symmetries and disorder correlations, we find that the universal conductivity of these models is approximately
the same. In particular, the ratio of this value to that of the  pure model agrees with recent experiments on nanohole thin film superconductors in a magnetic field,  in the large disorder limit. 

\end{abstract}

\maketitle

\section{Introduction}
Models of phase coherence in inhomogeneous superconductors, which incorporate gauge disorder, have been widely
used to study the vortex glass transition of type II superconductors driven by thermal fluctuations \cite{fisher89,huse90}. Gauge disorder appears as random phase shifts  in the Josephson junctions  coupling local superconducting islands, due to the combined effect of geometrical disorder and the  applied magnetic field. Phase shifts can also  arise from the presence of negative Josephson couplings or $\pi$ junctions \cite{bula77,spivak91,sigrist95}, even in the absence of the magnetic field, and can lead to different phase transitions and changes in the magnetic properties\cite{kusma92,kawa95,kawa97,eg98}. Although there
are many recent studies of the effects of disorder both in two dimensional  and one dimensional \cite{vojta2012,basko2013,rastelli2015}  systems,   
the superconductor to insulator (SI) transition described by the quantum version of random gauge models has been, to a certain extent, much less investigated \cite{phillips03,stroud08,tang08,eg16R}. The magnetic field induced SI transition in thin films has actually  been studied in detail using disordered Bose-Hubbard models \cite{fisher90a,fisher90b}, which include a random potential, but the additional effects of gauge disorder is difficult to be included in  the numerical simulations \cite{wallin94}. There are, however, interesting superconducting systems in the form of thin films with a pattern of nanoholes  \cite{valles07,valles08,batur11,kopnov,valles15,valles16} and micro-fabricated Josephson-junctions arrays \cite{vdzant,chen95,fazio,han2014collapse}, where gauge disorder alone  should play a dominant effect in the properties of the SI transition. Such systems allow comparisons with  the results from minimal random gauge models. 

Very recently \cite{valles15,valles16}, the effect of controlled amount of gauge disorder on the SI transition was investigated  
in nanohole ultrathin films by introducing geometrical disorder  in the form of randomness in the positions of the nanoholes. 
A minimal model describing phase coherence in these systems consists of a Josephson-junction array defined on an appropriate lattice, with the nanoholes corresponding to the dual lattice \cite{eg13,eg16,eg16R}. Positional disorder of the grains or in the plaquette areas \cite{gk86,gk86b,benz88,stroud08}, leads to disorder in  the magnetic flux per plaquette which increases with the applied field and geometrical disorder strength. Magnetoresistance oscillations near the SI transition, resulting  from commensurate vortex-lattice states, are observed below a critical  disorder strength \cite{valles15}. While the resistivity at
the successive field-induced transitions varies below this critical disorder, it reaches a constant value, independent of the critical coupling for larger 
disorder \cite{valles16}. Recent numerical simulations of a Josephson-junction array model suggest that the large disorder regime
should correspond to a vortex glass \cite{eg16R}.  Random gauge models with quantum fluctuations (quantum rotor models) should then provide the simplest description for the SI transition in this limit. Since the choice of the appropriate model is not unique, it should be of interest to compare the results for different models. 

In this work, we study numerically the SI transition in two-dimensional inhomogeneous superconductors described by random gauge models. Four different quantum rotor models are considered: a gauge glass, a flux glass,  a binary phase glass and a Gaussian phase glass. The first two models, describe the combined effect of geometrical disorder in the array of local superconducting islands and a uniform external magnetic field while the last two describe the effects of randomness in the Josephson couplings alone,  allowing for negative couplings. Monte Carlo simulations in the path-integral representation are used to determine the critical exponents and the electrical conductivity at the transition.  We find that the gauge and flux glass models display the same critical behavior, within the estimated numerical uncertainties. Similar agreement is found for the binary and Gaussian phase-glass models.  Despite the different symmetries and disorder correlations, the universal conductivity of these models is approximately the same. We compare the results for gauge and flux glass models with recent experiments on nanohole thin film superconductors in a magnetic field with controlled amount of gauge disorder \cite{valles15,valles16}. 
In particular, the ratio of the critical  conductivity for large gauge disorder to that of the pure model is in good agreement with the experimental data. The
results support the experimental observation \cite{valles16} that the critical conductivity is independent of the coupling constant for large disorder, consistent with the scenario of a universal value in this limit.  

\section{Models and Monte Carlo simulation}

We consider models which describe  two-dimensional superconductors as an array of Josephson junctions, allowing for charging effects  and gauge disorder \cite{bradley84,gk86,stroud08,fazio,eg16R}, defined by the Hamiltonian
\begin{equation}
{\cal H} = {{E_C}\over 2} \sum_i n_i^2 - \sum_{<ij>} E_{ij} \cos ( \theta_i -
\theta_{j}-A_{ij}). \label{hamilt}
\end{equation}
The first term in Eq. (\ref{hamilt}) describes quantum
fluctuations induced by the charging energy, $ E_C  n_i ^2/ 2$,  of a
non-neutral superconducting "grain", or "island", located at site $i$ of a reference square lattice, where $E_C= 4 e^2/C$,
$e$ is the electronic charge, and
$n_i= -i \partial /\partial \theta_i $ is the operator, canonically conjugate to the phase operator $\theta_i$, representing the deviation of the number of Cooper pairs from
a constant integer value. The effective capacitance to the ground of each grain $C$
is assumed to be spatially uniform, for simplicity.
%A more realistic model allowing an intergrain
%capacitance should not change the universality class, which depends on
%the on small wavevector behavior of the capacitance matrix \cite{fisher88}.
The second term in (\ref{hamilt}) is the
Josephson-junction coupling between nearest-neighbor grains
described by phase variables  $\theta_i$ and phase shifts $A_{ij}$. 
%For a  spatially uniform Josephson coupling,  $E_{ij}=E_J$. 
The model in Eq. (\ref{hamilt}) can also be regarded as a quantum rotor model \cite{cha3} with the additional effects of quenched gauge disorder. 
We consider four different quantum rotor models: a gauge and a flux-glass model  
\cite{gk86,gk86b,stroud08,eg16R} with spatially randomness in 
$A_{ij}$  and a  binary and Gaussian phase-glass model \cite{phillips03}  with  spatially randomness in  $E_{ij}$  including  $E_{ij} < 0 $. The phase-glass 
model can also be regarded as a quantum version of the chiral-glass model \cite{kawa95,weng97, eg04} used to study the thermal phase transition, in absence of charging effects. 

For the gauge and flux glass models, $A_{ij}$  represents  the line integral of the vector potential ${ A}_{ij}=\frac{2 \pi}{\Phi_o} \int_i^j{\bf A} \cdot d {\bf l} $, due to an external magnetic field ${\bf B = \nabla \times \bf A}$.  
%The sum of $A_{ij}$ around a plaquette $p$ of area $S_p$ can be written as $sum_{ij} A_{ij}=2 \pi f_p$, where $f_p=B \delta S_p/\Phi_o $ is the magnetic flux in units of the flux quantum $\Phi_o=hc/2e$.
For the {\it gauge-glass model}, we set $E_{ij}=E_J$ (uniform) and choose  $A_{ij}$ as a random variable uniformly distributed in the interval $[-\pi,\pi]$ but uncorrelated in space. It may describe, for example, the limit of very large disorder in the positions of the superconducting grains. 
%In the {\it flux-glass model}, $f_p$ is the spatially uncorrelated random variable, which we choose to be uniform in the interval interval $[-1,1]$.
%This could represent  a large disorder in the size of the grains, which induces uncorrelated variations in the magnetic flux at different plaquettes or 
%randomness in the plaquette areas $\delta S_p$, which leads to variations in the magnetic flux  $\delta f_p$. 
In the {\it flux-glass model}, the variation of the magnetic flux  $\delta f_p=B \delta S_p/\Phi_o$  in a plaquette  of area $S_p$, in units of the flux quantum  $\Phi_o=hc/2e$, is the spatially uncorrelated random variable, which we choose to be uniform in the interval interval $[-1,1]$. This could represent  a large disorder in the size of the grains, which induces uncorrelated variations in the magnetic flux at different plaquettes or 
randomness in the plaquette areas.  The flux-glass model can also be regarded as a gauge-glass model with a particular long-range correlated disorder \cite{gk86,gk86b} in $A_{ij}$. The {\it phase-glass model} describes the effects of disorder in $E_{ij}$ due to random location of negative Josephson coupling ($E_{ij}<0$). In this case, we set $A_{ij}=0$ and choose 
%$E_{ij}$ to be Gaussian distributed with probability $P(E_{ij})=\exp(-E_{ij}^2/2E_J)/\sqrt{2\pi} $.
$E_{ij} = \pm E_J$, with equal probability (binary distribution)  or with probability $P(E_{ij}) =e^{-E_{ij}^2/2 E_J^2 }/E_J \sqrt{2 \pi}$ (Gaussian distribution).  Since $E_{ij}<0$ with $A_{ij}=0$ is equivalent to a positive Josephson coupling 
$|E_{ij}|$ with a phase shift $A_{ij} = \pi $, the binary phase-glass model can also be regarded as a gauge-glass model with a binary distribution of phase shifts $A_{ij}=0$ or $ \pi $.

The quantum phase transition at zero temperature can be conveniently studied in the framework of the  imaginary-time path-integral formulation of the model \cite{sondhi}. In this representation, the two-dimensional (2D) quantum model of Eq. (\ref{hamilt}) maps into a (2+1)D classical statistical mechanics problem. The extra dimension corresponds to the
imaginary-time direction. Dividing the time axis $\tau$ into slices $\Delta \tau$, the ground state energy corresponds to the reduced free energy $F$ of the classical model per time slice. The classical reduced Hamiltonian can be written as \cite{bradley84,wallin94,sondhi}
\begin{eqnarray}
H= &&-\frac{1}{g} [ \sum_{\tau,i}
\cos(\theta_{\tau,i}-\theta_{\tau+1,i}) \cr &&
+\sum_{<ij>,\tau} e_{ij}\cos(\theta_{\tau,i}-\theta_{\tau,j}-A_{ij}) ],
\label{chamilt}
\end{eqnarray}
%where a re-scaling of the time slices has been performed in order to obtain space-time isotropic couplings and
where $ e_{ij}=E_{ij}/ E_J$ and $\tau$ labels the sites in the discrete time direction. 
The ratio $g =(E_C/E_J)^{1/2}$, which drives the SI transition for the
model of Eq. (\ref{hamilt}), corresponds to an effective "temperature" in the
3D classical model of Eq. (\ref{chamilt}). The  particular form of the coupling of the phases $\theta_{\tau,j}$ 
in the time direction results from a Villain approximation, used  to obtain the phase representation
of the first term in Eq. (\ref{hamilt}). This approximation, however,  should preserve  the universal aspects of the critical behavior \cite{sondhi}. 
In general, a quantum phase transition shows intrinsic anisotropic scaling, with  different diverging correlation
lengths $\xi$ and $\xi_\tau$ in the spatial and imaginary-time directions, respectively, related by the dynamic critical exponent $z$
as $\xi_\tau \propto \xi^z$.
%The energy gap $\Delta$ of the insulating phase is related
%to the phase correlation length in the time direction $\xi_\tau$ by $\Delta=1/\xi_\tau$.
The classical Hamiltonian of Eq. (\ref{chamilt}) can be viewed as a three-dimensional (3D)  layered 
XY model , where  frustration effects exist
only in the 2D layers. Randomness in $e_{ij}$ or  $A_{ij}$ corresponds to disorder
completely correlated in the time direction.

Equilibrium Monte Carlo (MC) simulations  are carried out using the
3D classical Hamiltonian in Eq. (\ref{chamilt}) regarding $g$ as a "temperature"-like parameter.
The parallel tempering method \cite{nemoto} is used in the simulations
with periodic boundary conditions, as in previous work \cite{eg16,eg16R}.
%Since the correlation lengths in the spatial and imaginary-time directions are related by dynamical scaling
%as $\xi_\tau \propto \xi^z$,
The finite-size scaling analysis is performed for different linear sizes $L$ of the square lattice with the constraint
$ L_\tau =a L^z $, where $a$ is a constant aspect ratio. This choice simplifies the scaling analysis, otherwise
an additional scaling variable $L_\tau/L^z$ would be required to describe the scaling functions.
The value of $a$ is chosen to minimize the deviations of $ a L^z$ from integer numbers. However, this requires one to know the value of the dynamic exponent $z$ in advance.
%However, in presence of disorder, to estimate the value of $z$  with the above {\it equilibrium} MC method is %computationally very demanding. It requires performing simulations and disorder averages for increasing system sizes %$L_\tau$ in the time direction to find the optimum value of
%$z$ that gives the best scaling behavior.
%To overcome this problem, we follow a two-step approach.
Since the exact value of $z$ is not known, we follow a two-step approach.
First, we obtain an estimate of $g_c$ and $z$ from  simulations performed with a {\it driven} MC dynamics method, which has been used in the context of the 3D classical XY-spin glass model \cite{eg04}. Then, these initial estimates are improved by finding the best data collapse for the finite-size behavior of the phase stiffness in the time direction $\gamma_\tau$, obtained by the equilibrium MC method. 

For the driven MC method, the layered honeycomb model of Eq. (\ref{chamilt}) is viewed as a 3D superconductor and the corresponding "current-voltage" scaling near the transition is used to determine the critical coupling and critical exponents \cite{weng97}. In the presence of an external driving perturbation $J_x$ ("current density") which couples to the phase difference $\theta_{\tau,i + \hat x}-\theta_{\tau,i}$ along the $\hat x$ direction, the classical Hamiltonian of Eq. (\ref{chamilt}) is modified to
\begin{eqnarray}
H_J= H -\sum_{i,\tau} \frac{J_x}{g} (\theta_{\tau,i+\hat x}-\theta_{\tau,i}).
\label{driven}
\end{eqnarray}
The MC simulations are carried out using the Metropolis algorithm and
the time dependence is obtained from the MC time $t_{mc}$.
When $J_x \ne 0$,  the system is out of equilibrium since the  total energy is unbounded.
The lower-energy minima occur at phase differences $\theta_{\tau,i+\hat x}-\theta_{\tau,i}$,
which increase with time $t_{mc}$, leading to a net phase slippage
rate proportional to $ V_x = <d(\theta_{\tau,i+\hat x}-\theta_{\tau,i})/dt_{mc} > $, corresponding to the average
"voltage" per unit length.  The measurable quantity
of interest is the phase slippage response ("nonlinear resistivity") defined as $R_x = V_x/J_x$.  Similarly,
we define $R_\tau$ as the phase slippage response to the applied perturbation $J_\tau$ in the layered (imaginary-time) direction.
Above the phase-coherence
transition, $g > g_c$, $R_x$ should approach a nonzero value when $J_x \rightarrow 0$ while it should  approach zero below the transition. From the nonlinear scaling behavior near the transition of a sufficiently large system, one can extract the critical coupling $g_c$, and the critical exponents $\nu$ and $z$.
%In the absence of charging effects, $R_x$ remains zero below a critical value $J_x=J_c$, which provides an
%estimate of the critical current for  the model of Eq. (\ref{hamilt}), when $E_c=0$.
%The scaling theory describing this behavior has been developed in the context of the current-voltage characteristics of %vortex-glass models \cite{vortexg}. In the present case, it needs to be generalized to take into account anisotropic %scaling, where $z\ne 1$. The required generalization has been described in detail in Ref. \onlinecite{girvin}.

\section{Numerical results and discussion}
\begin{figure}
\centering
\includegraphics[width=\columnwidth]{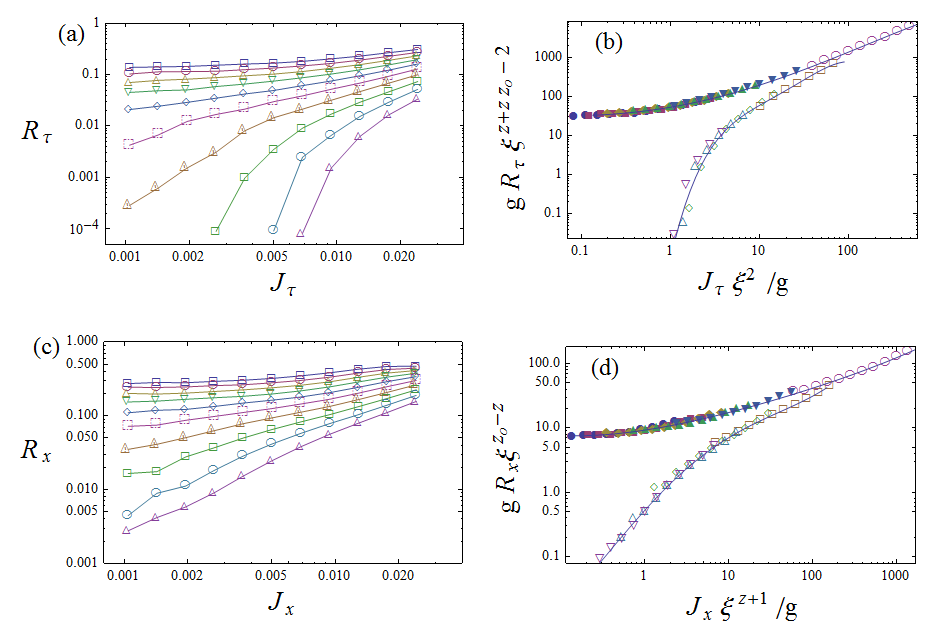}
\caption{ Scaling behavior of the  phase slippage response for the gauge-glass model in (a) the imaginary-time direction $R_\tau$ and (c) spatial direction $R_x$ near the SI
transition. From the top down, the couplings are $g=1.75, 1.73, 1.71, 1.69, 1.67, 1.65, 1.63, 1.61, 1.59 $  and $ 1.57 $.
(b) and (d) Scaling plots corresponding to (a) and (c), respectively, for data near the transition with $\xi=|g/g_c-1|^{-\nu}$ using the same 
parameters $g_c=1.645$, $ z_o=2.3$, $z=1.2$, and $\nu = 0.9$. }
\label{gcgg}
\end{figure}

A first estimate of the critical coupling $g_c$ and dynamical exponent $z$ can be obtained using the driven MC dynamics method presented in Sec. II for large system sizes. We illustrate the method for the gauge-glass model.  Figs. \ref{gcgg} shows the behavior of the nonlinear phase slippage response $R_x$ and $R_\tau$ for the gauge-glass model as a function of the applied perturbation $J_x$ and $J_\tau$, respectively. The behavior for different values of $g$ is consistent with a phase-coherence transition at an apparent
critical coupling in the range $g_c \sim 1.63 - 1.67$. For $g > g_c$, both
$R_x$ and $R_\tau$ tend to a finite value while for $g < g_c$, they
extrapolate to low values. Assuming the transition is continuous,
the nonlinear response behavior sufficiently close the transition should satisfy a scaling form in terms of $J_x$, $J_\tau$ and $g$.
The critical coupling $g_c$ and critical exponents $\nu$ and $z$ can then be obtained from the
best data collapse satisfying the scaling behavior close to the transition. Details of the scaling theory can be found in ref. \onlinecite{girvin}.
$R_x$  and $R_\tau$ should satisfy the scaling forms
\begin{eqnarray}
g R_x \xi^{z_0-z} & = & F_\pm (J_x \xi^{z+1}/g),  \cr
g R_\tau \xi^{z + z_0 z - 2} & = & H_\pm (J_\tau \xi^2/g),
\label{Rxy}
\end{eqnarray}
where $z_o$ is an additional critical exponent describing the MC relaxation times, $t^r_{mc, x}\sim \xi^{z_o}$ and  $t^r_{mc,\tau}\sim \xi_\tau^{z_o}$, in the spatial and imaginary-time directions, respectively,  and  $\xi=|g/g_c-1|^{-\nu}$.
The + and -  signs correspond to $g  > g_c$ and $g  < g_c$, respectively. The two scaling forms are the same when $z=1$, corresponding to isotropic scaling.  The joint scaling plots according to Eqs. (\ref{Rxy}) are shown in  Figs. \ref{gcgg}b and  \ref{gcgg}d obtained by adjusting the unknown parameters, providing the estimates $g_c=1.645 $, $z_o=2.3 $, $z=1.2 $ and $\nu=0.9$. 

\begin{figure}
\centering
\includegraphics[width=\columnwidth]{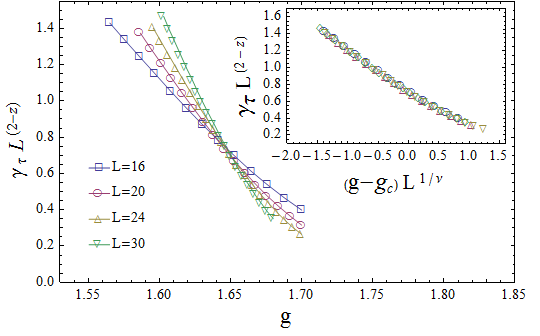}
\caption{  Phase stiffness in the imaginary time direction $\gamma_\tau $ for the gauge-glass model with 
different system sizes $L$, near the transition point estimated from Figs. \ref{gcgg}. $L_\tau=a L^z$,
with aspect ratio $a=0.642$ and $z=1.2$.
Inset: scaling plot of $\gamma_\tau $ with  $g_c= 1.649 $  and $\nu = 0.99$. .}
\label{gg}
\end{figure}

\begin{figure}
\centering
\includegraphics[width=0.9\columnwidth]{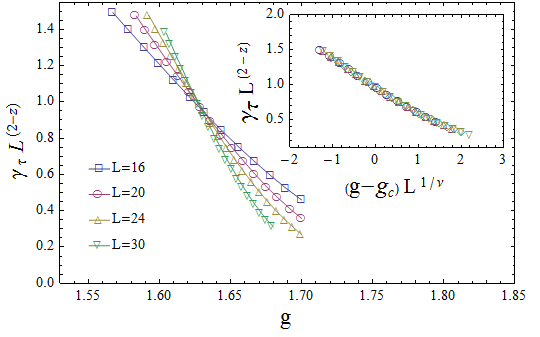}
\caption{ Same as Fig. \ref{gg} but for the flux-glass model.  
Inset: scaling plot of $\gamma_\tau $ with $g_c =1.6294 $ and $ \nu =0.92 $ .}
\label{fluxg}
\end{figure}

\begin{figure}
\centering
\includegraphics[width=0.9\columnwidth]{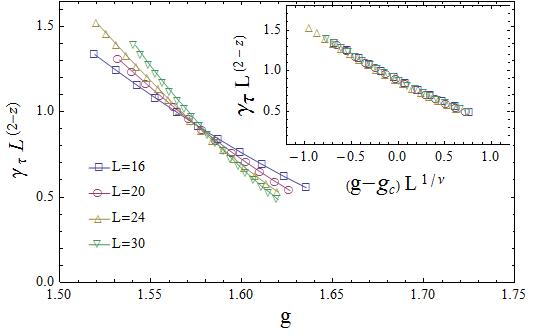}
\caption{ Same as Fig. \ref{gg} but for the phase-glass model.  
Inset: scaling plot of $\gamma_\tau $ with $g_c =1.58 $ and $ \nu =1.15 $ .}
\label{cg}
\end{figure}

To obtain the estimates above, it was implicitly assumed that he system is sufficient large and the coupling parameter is not too close to $g_c$, allowing the finite-size effects to be neglected.
Having obtained an estimate of $z$, we can now  consider the  finite-size behavior of the phase stiffness in the imaginary time direction $\gamma_\tau$, using equilibrium MC simulations,
and improve the determination of $g_c$ and $\nu$.  The phase stiffness $\gamma_\tau$, which is
a measure of the free energy cost to impose an infinitesimal phase twist in the time
direction, is given by \cite{cha2}
\begin{eqnarray}
\gamma_\tau=\frac{1}{L^2 L_\tau g^2}[g <\epsilon_\tau>
- < I_\tau^2 > + < I_\tau >^2]_D,
\label{eqstiff}
\end{eqnarray}
where $\epsilon_\tau = \sum_{\tau,i} \cos(\theta_{\tau,i} -\theta_{\tau+1,i})$ and   $I_\tau= \sum_{\tau,i}\sin(\theta_{\tau,i} -\theta_{\tau+1,i})$. In Eq. (\ref{eqstiff}), $< \ldots>$ represents a MC average
for a fixed disorder configuration and $[  \ldots ]_D$ represents an average over different disorder configurations.
In the superconducting phase $\gamma_\tau$ should be finite, reflecting the existence of phase coherence, while in the
insulating phase it should vanish in the thermodynamic limit.
For a continuous phase transition, $\gamma_\tau$ should satisfy the finite-size scaling form
\begin{equation}
\gamma_\tau L^{2-z} = F(L^{1/\nu} \delta g),
\label{rhotau}
\end{equation}
where $F(x)$ is a scaling function and $\delta g = g-g_c$. This scaling form implies that data for $ \gamma_\tau L^{2-z} $ as a function of $g$, for different system sizes $L$, should cross at the critical coupling $g_c$. Fig. \ref{gg}a shows this crossing behavior obtained near the initial estimate of $g_c$ obtained from Fig. \ref{gcgg}. by varying slightly $g_c$ and $\nu$  from their initial values. In the Inset of this Figure, we show a scaling plot of the data according to the scaling form of Eq. \ref{rhotau}, which provides for the gauge-glass model the final estimates $g_c= 1.649 $  and $\nu = 0.99$.  The same value of the dynamic exponent $z=1.2$ found for the gauge-glass model also give consistent results for the other models.  Figures   \ref{fluxg} and \ref{cg} show the scaling behavior of the phase stiffness for the flux and binary phase-glass model.  We then obtain the estimates $g_c=1.629 $ and $\nu=0.92 $ (flux glass),  $g_c=1.58 $ and $\nu=1.15 $ (binary phase glass),   $g_c=1.44$ and $\nu=1.12$ (Gaussian phase glass). 

\begin{figure}
\centering
\includegraphics[width=0.9\columnwidth]{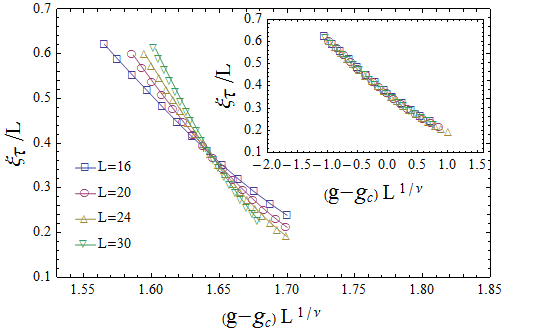}
\caption{  Correlation length in the imaginary time direction $\xi_\tau $ for the gauge-glass model with 
different system sizes $L$. Inset: scaling plot of $\xi_\tau $ with  $g_c= 1.646 $  and $\nu = 1.08$. .}
\label{taugg}
\end{figure}

\begin{figure}
\centering
\includegraphics[width=0.9\columnwidth]{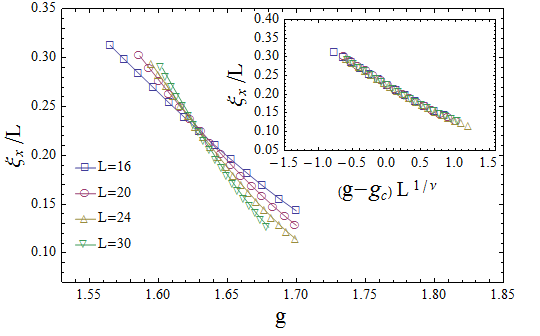}
\caption{  Correlation length in the spatial direction $\xi $ for the gauge-glass model with 
different system sizes $L$. Inset: scaling plot of $\xi $ with  $g_c= 1.629 $  and $\nu = 1.12$. .}
\label{xgg}
\end{figure}

The SI transition can be further characterized by the  behavior of the finite-size correlation length, which can be defined as \cite{balle00}
\begin{equation}
\xi(L,g) = \frac{1}{2\sin(k_0/2)}[S(0)/S(k_0) - 1]^{1/2}.
\label{cdef}
\end{equation}
Here $S(k)$ is  the Fourier transform of the correlation function $C(r)$ and $k_0$ is the smallest nonzero
wave vector. For $g > g_c $, this definition  corresponds to  a finite-difference approximation to
the infinite system correlation length
$\xi^(g)^2= -\frac{1}{S(k)} \frac{\partial S(k) }{\partial k^2} |_{k=0}$,
taking into account the lattice periodicity. For the random-gauge models considered here, it is convenient to define the correlation function in
terms of the overlap order parameter \cite{bhat88} $q_{\tau,j}=\exp( i(\theta^1_{\tau, j} - \theta^2_{\tau,j}) )$, 
where $1$ and $2$ label two different copies of the system with the same coupling parameters. The correlation function in the spatial direction
is obtained as 
\begin{equation}
C(r) =\frac{1}{L^2L_\tau} \sum_{\tau,j} <q_{\tau,j} q_{\tau,j+r}>,
\label{cfunc}
\end{equation}
and the analogous expression is used for the correlation function $C_\tau(r)$ in the time direction.
For a continuous transition, $\xi(L,g)$ should satisfy the scaling form 
\begin{equation}
\xi/L = F(L^{1/\nu} \delta g ),
\label{correl}
\end{equation}
where $F(x)$ is a scaling function.
Figures  \ref{taugg} and \ref{xgg} show the behavior of the correlation length  $\xi_\tau $  and  $\xi$ in the time and spatial directions, for the gauge-glass model. The curves  for $\xi_\tau/L $ as a function of $g$ for different system sizes cross at the same point, providing further evidence of a continuous transition. In the inset  of  Fig. \ref{taugg}, a scaling plot according to Eq. (\ref{correl}) is shown, which gives  an alternative estimate of $g_c=1.646$ and $\nu=1.08$.  For the correlation length in the spatial direction shown in Fig. \ref{xgg} and  the corresponding scaling plot, we obtain $g_c=1.629$ and $\nu=1.12$.  Since in this case the crossing point is less clear, these estimates are more affected by corrections to finite-size scaling. For the flux and phase-glass models the difference of the estimate of $g_c$ from the correlation in the time and spatial directions are much larger.  We consider that the results obtained from the scaling of the phase stiffness $\gamma_\tau$ 
% and correlation length $\xi_\tau$
are more accurate and use them to obtain the final result and the associated errorbar. 

\begin{figure}
\centering
\includegraphics[width=1.0\columnwidth]{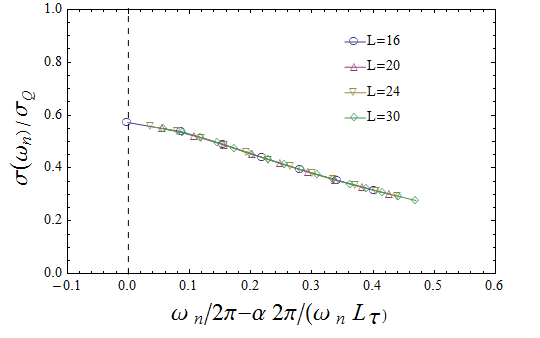}
\caption{  Scaling plot of conductivity $\sigma(iw_n)$ at the critical coupling $g_c$ for the gauge-glass model
with $\alpha=0.2$. The universal conductivity  is given by the intercept with the $x=0$
dashed line, leading to  $\frac{\sigma^*}{\sigma_Q}= 0.56(3)$. }
\label{condgg}
\end{figure}

\begin{figure}
\centering
\includegraphics[width=1.0\columnwidth]{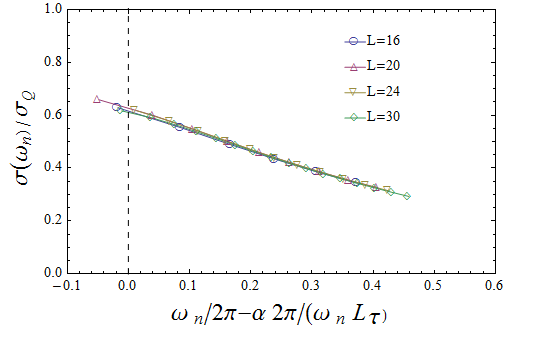}
\caption{ Same as Fig. \ref{condgg} but for the flux-glass model with $\alpha=0.27$.  The universal conductivity  is given by the intercept with the $x=0$
dashed line, leading to  $\frac{\sigma^*}{\sigma_Q}= 0.61(3)$. }
\label{condfluxg}
\end{figure}

\begin{figure}
\centering
\includegraphics[width=1.0\columnwidth]{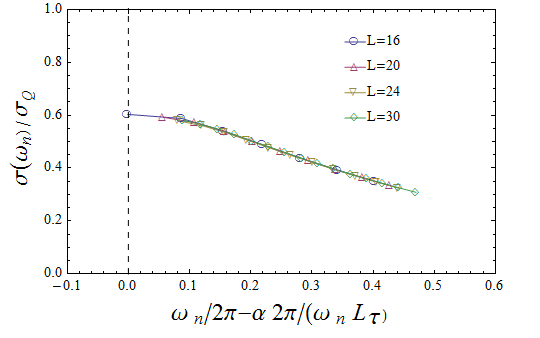}
\caption{ Same as Fig. \ref{condgg} but for the binary phase-glass model with $\alpha=0.06$.  The universal conductivity  is given by the intercept with the $x=0$
dashed line, leading to  $\frac{\sigma^*}{\sigma_Q}= 0.60(3)$. }
\label{condcg}
\end{figure}

We have also determined the universal conductivity at the critical point from the frequency and finite-size dependence of the phase stiffness $\gamma(w)$ in the spatial direction, following the scaling method described by Cha {et al.} \cite{cha2,cha3}.
The conductivity is given by the Kubo formula
\begin{equation}
\sigma = 2 \pi \sigma_Q \lim_{w_n\rightarrow 0}  \frac{\gamma(i w_n)}{w_n},
\end{equation}
where $\sigma_Q=(2 e)^2/h$ is the quantum of conductance and $\gamma(i w_n)$ is a frequency dependent phase stiffness
evaluated at the finite frequency $w_n=2 \pi n /L_\tau$, with $n$ an integer. The frequency dependent  phase stiffness in
the $\hat x$ direction is given by
\begin{eqnarray}
\gamma=\frac{1}{L^2 L_\tau g^2} [& g &< \epsilon_x >
-< |I(i w_n)|^2 > \cr
& + & < |I(i w_n)| >^2]_D,
\end{eqnarray}
where
\begin{eqnarray}
\epsilon_x&=&\sum_{\tau,j}  e_{i,j+\hat x}\cos(\Delta_x \theta_{\tau,j}) ,\cr
I(i w_n)&=&\sum_{\tau,j}  e_{i,j+\hat x} \sin(\Delta_x \theta_{\tau,j}) e^{i w_n \tau},
\end{eqnarray}
and $\Delta_x \theta_{\tau,j}= \theta_{\tau,j}-\theta_{\tau,j+\hat x} - A_{j,j+\hat x}$.
At the transition, $\gamma(i w_n)$ vanishes linearly with frequency and $\sigma$ assumes a universal value  $\sigma^*$, which can be  extracted from its frequency and finite-size dependence as  \cite{cha2}
\begin{equation}
\frac{\sigma(iw_n)}{\sigma_Q} = \frac{\sigma*}{\sigma_Q}
- c (\frac{w_n}{2 \pi} - \alpha \frac{2 \pi}{w_n L_\tau}) \cdots
\label{cond}
\end{equation}
The parameter $\alpha$ is determined from  the best data collapse of the frequency
dependent curves for  different systems sizes  in a plot of $\frac{\sigma(iw_n)}{\sigma_Q}$ versus
$x=(\frac{w_n}{2\pi} - \alpha \frac{2\pi}{w_n L_\tau})$. The universal conductivity is
obtained from the intercept of these curves with the line $x=0$.
The calculations were performed for different system sizes with $L_\tau=a L^z$, using the above estimates of $z$ and $g_c$. From the scaling behavior in Fig. \ref{condgg} we obtain  for the gauge-glass model $\sigma^*/\sigma_Q = 0.56(3)$, where the estimated uncertainly is mainly the result of the error in the coupling $g_c$.  Fig. \ref{condfluxg} and Fig. \ref{condcg} show the behavior for the flux and binary phase-glass models. We then obtain   $\sigma^*/\sigma_Q = 0.61(3)$ (flux glass),  $\sigma^*/\sigma_Q = 0.60(3)$ (binary phase glass)
and $\sigma^*/\sigma_Q = 0.57(3)$ (Gaussian phase glass)

\begin{table}[h]
\begin{center}

\begin{tabular}{ c c c c c c}
%\begin{tabular}{ c|c|c|c|c|c}
 \hline
     & pure  &   gauge       &     flux  &    binary       &    Gaussian  \\
    &          &    glass     &    glass  &  phase glass &   phase glass \\
 \hline
  $g_c$ & $2.203$  & $1.649(1) $  & $1.629(1)$  & $1.580(5) $ & $1.440(5) $  \\
  \hline
  $z$   & $1$  & $1.2(1)$ & $1.2(1) $ & $1.2(1)$ & $1.2(1)$ \\
 \hline
 $\nu$ & $0.67$  &  $0.99(4)$ & $0.92(4)$   & $1.15(6)$ & $ 1.12(4)$ \\
 \hline
 $\sigma^*/\sigma_Q$ & $ 0.29(2)$ & $0.56(3)$ &  $0.61(3)$ & $0.60(3)$ & $0.57(3)$ \\
 \hline
 \end{tabular}
 \caption{\label{tab:disor} Critical  exponents $z, \nu $ and critical conductivity $\sigma^*$ for different random gauge models and
the pure model (without disorder). $g_c$ is the critical value of the coupling parameter $g=(E_C/E_J)^{1/2}$. The results for the pure case are taken
from ref. \onlinecite{cha3}. }

\end{center}
\end{table}

The results for the critical properties of  the different random gauge models are summarized in Table I, together with
the known values for the pure model.  We now compare them with available numerical work and  experimental data. 
The value of the universal conductivity found in the earlier work on the gauge-glass model \cite{stroud08}, $\frac{\sigma^*}{\sigma_Q}= 1.06(9)$, differs significantly from our result but the critical exponent $z=1.3(1)$ is consistent with our estimate. The discrepancy in the value of $\sigma^*$ is mainly due to  the different estimate of the critical coupling, $g_c\approx 1.587$, which was obtained by a scaling analysis of the dimensionless ratio of the overlap order parameter $q_{\tau,j}$. This type of "Binder ratio", however, is  not very reliable for models with continuous symmetry \cite{shira03}. Since our estimate of $g_c$ is based on  the scaling behavior of the phase stiffness, which is also consistent with the behavior of the correlation length, we believe it should be more accurate. A different calculation of the critical exponents \cite{tang08} found $ \nu\approx 0.73$ and  estimate $z=1.17(7)$,  also compatible with our result for $z$.  

The results for the gauge and flux-glass models can be compared with experimental observations of the SI transition on thin superconducting films with a pattern of nanoholes \cite{valles07,valles15,valles16}. A minimal model describing phase coherence in these systems consists of a Josephson-junction array defined on an appropriate lattice, with the nanoholes corresponding to the dual lattice \cite{eg13,eg16,eg16R}.  Very recently \cite{valles15,valles16},
the effect of controlled amount of gauge disorder on the SI transition was investigated  by introducing geometrical disorder 
in the form of randomness in the position of the nanoholes. This leads to disorder in the magnetic flux  
$ \delta f_p=B \delta S_p/\Phi_o$ in a plaquette of area  $S_p$, which increases with the applied magnetic field and degree of geometrical disorder.  Magnetoresistance oscillations near the SI transition, resulting  from commensurate vortex-lattice states, are observed below a critical  disorder strength \cite{valles15} $\delta f_c \approx 0.3 $. Although the resistivity at successive field-induced SI transitions varies below this critical disorder, it seems to reach 
a constant value, independent of the critical coupling for larger disorder \cite{valles16}. Recent numerical simulations of a Josephson-junction array model suggests that the large disorder regime should correspond to a vortex glass \cite{eg16R}. The gauge  and flux-glass models considered here should
then provided the simplest description in this limit. For weak geometrical disorder, the nanoholes form a triangular lattice \cite{valles07} and therefore the appropriate geometry for the array model should be a honeycomb lattice \cite{eg16R,eg17conf}.  In the large disorder limit, however, the lattice geometry should not be relevant.  In fact, the numerical results for the conductivity at the transition found for a flux-glass model using a honeycomb lattice in the large disorder limit \cite{eg16R} is the same, within the estimated errorbar, as found in the present work for the square lattice.  In particular,  the value of conductivity at the transition found in the experiments for large gauge disorder \cite{valles15,valles16} is a factor of $~2$ larger compared with measurements on samples without an applied magnetic field \cite{valles07}. This ratio of the critical conductivities  agrees with the results for the gauge or flux-glass models compared with the pure model in Table I.  Therefore, although the magnitudes of the experimental and numerical results are different, the trend of increasing critical conductivity with gauge disorder is correctly given by the gauge and flux-glass models. Notice, however, that the opposite trend can occur when comparing the large gauge disorder limit  with the pure system in presence of a magnetic field \cite{eg16R}.  Moreover, the agreement of the critical properties obtained from the gauge and flux-glass models and the previous calculations for large disorder from  a model on a honeycomb lattice \cite{eg16R}, strongly supports the experimental observation \cite{valles16} that the critical conductivity is independent of the coupling parameter in the large disorder limit. 

It may appear somehow surprising that the critical conductivity for the phase-glass model is essentially the same as for the gauge-glass model. The phase-glass model has an additional reflection symmetry property \cite{kawa95}, where changing $\theta_i \rightarrow -\theta_i$ leaves the Hamiltonian unchanged, whereas for the gauge-glass model there is only a continuous symmetry. One could then expect different universality classes.  In the absence of quantum fluctuations, $E_c=0$, this happens to be the case. In 2D, the transition for increasing temperatures can be described as a thermal transition with vanishing critical temperature, $T_c=0$, and a divergent  thermal correlation length $\xi_T \propto T^{-\nu_T}$.  In fact, the value of $\nu_T$  for the gauge and phase-glass models are quite different \cite{kawa91,eg98}.  On the other hand, the SI transition at zero temperature is actually described by an effective $(2 +1)$D classical model (Eq. (\ref{chamilt})) with gauge disorder completely correlated in one direction.  Interestingly, numerical results for the 3D gauge and phase-glass models show the same critical exponents \cite{weng97,eg04,leeyoung}, within the estimated errorbar, although such calculations have only been carried out for models with uncorrelated disorder. 

%\acknowledgements

\section{Conclusions}
We studied the superconductor-insulator transition in two-dimensional inhomogeneous superconductors with
gauge disorder, described by four different models: a gauge glass, a flux glass, a binary phase glass and Gaussian phase-glass model. The first two models, describe the combined effect of geometrical disorder in the array of local superconducting islands and a uniform external magnetic field while the last two describe the effects of randomness in the Josephson couplings alone,  allowing for negative couplings.   We found that the gauge and flux-glass models display the same critical behavior, within the estimated uncertainties, and similar behavior is observed for binary and Gaussian phase-glass models.  The value of the conductivity at the transition is a factor of 2 larger than for the  pure model, which agrees with recent experiments on nanohole thin film superconductors \cite{valles15,valles16} in the large disorder limit, which can be modeled by the gauge or flux-glass models.  This agreement together with previous results for large disorder from  a model on a honeycomb lattice \cite{eg16R}, strongly supports the experimental observation \cite{valles16} that the critical conductivity is independent of the coupling parameter in the large disorder limit,  consistent with the scenario of a universal value in this limit.  
For a more realistic description of these systems  dissipation effects \cite{fazio}, which have been neglected in the present
models, should also be taken into account.
It should be noted that the phase-glass models considered here, which show a direct superconductor to insulator transition, have a zero mean distribution of Josephson couplings. For a nonzero mean, an analytical work \cite{phillips03} has proposed an intermediate metallic phase (a Bose metal) separating the superconducting and insulating phases. 

\acknowledgements
The author thanks J. M. Valles Jr. and J.M. Kosterlitz for helpful discussions.
This work was supported by  CNPq (Conselho Nacional de Desenvolvimento Cient\'ifico e Tecnol\'ogico) in Brazil and computer facilities from CENAPAD-SP.

\bibliography{qglass}

\end{document}